\documentclass[showpacs,aps,prl,twocolumn,floatfix,superscriptaddress]{revtex4}
\usepackage{CJK}
\usepackage{graphicx}
\usepackage{epsfig}
\usepackage{wasysym}
\newcommand{\sumnn}{\mathop{\sum}_{\langle i, j \rangle_{\wedge}}}
\newcommand{\sumnear}{\mathop{\sum}_{\langle i, j \rangle_x}}

\begin{document}
\begin{CJK*}{UTF8}{gbsn}
\title{Quantum domain walls induce incommensurate supersolid phase \\ on the anisotropic triangular lattice}
\author{Xue-Feng Zhang (﻿张学锋)}
\thanks{Corresponding author: xuefeng@pks.mpg.de}
\affiliation{Physics Dept.~and Res.~Center
OPTIMAS, Univ.~of Kaiserslautern, 67663 Kaiserslautern, Germany}
\affiliation{Max-Planck-Institute for the Physics of Complex Systems, 
01187 Dresden}
\affiliation{State Key Laboratory of Theoretical Physics, Institute of Theoretical Physics, Chinese Academy of Sciences, Beijing 100190, China}
\author{Shijie Hu (胡时杰)}
\thanks{Corresponding author: shijiehu@physik.uni-kl.de}
\affiliation{Physics Dept.~and Res.~Center OPTIMAS, Univ.~of
Kaiserslautern, 67663 Kaiserslautern, Germany}
\author{Axel Pelster}
\affiliation{Physics Dept.~and Res.~Center OPTIMAS, Univ.~of
Kaiserslautern, 67663 Kaiserslautern, Germany}
\author{Sebastian Eggert}
\affiliation{Physics Dept.~and Res.~Center OPTIMAS, Univ.~of
Kaiserslautern, 67663 Kaiserslautern, Germany}
\date{\today}

\begin{abstract}
We investigate the extended hard-core Bose-Hubbard model on the
triangular lattice as a function of spatial anisotropy
with respect to both hopping and nearest-neighbor interaction strength.
At half-filling the system can be tuned from decoupled one-dimensional chains to a
two-dimensional solid phase
with alternating density order by adjusting the anisotropic coupling.
At intermediate anisotropy, however, frustration effects dominate and an
incommensurate supersolid phase emerges, which is characterized by incommensurate density order as well as an anisotropic
superfluid density.
We demonstrate that this intermediate phase
results from the proliferation of topological defects in the form of
quantum bosonic domain walls. Accordingly, the structure factor has peaks at wave
vectors, which are linearly related to the number of domain walls in a finite
system in agreement with
extensive quantum Monte Carlo simulations.
We discuss possible connections with the supersolid behavior in the high-temperature
superconducting striped phase.
\end{abstract}


\maketitle
\end{CJK*}

Ever since the observation of the superfluid-Mott transition in an optical atomic lattice
\cite{greiner},
ultra-cold gases have been
considered as promising candidates for the controlled quantum simulation
of condensed matter systems with interesting many-body physics \cite{bloch}.
Even though theorists are quite creative in inventing new models,
experimental progress is surprisingly quick to follow.
Recently, frustrated lattices \cite{fru} have been of broad
interest due to the emergence of new
exotic phases, such as
spin-liquids \cite{balents10,spinliquid1,spinliquid2,spinliquidn1,spinliquidn2,spinliquid3},
topological excitations \cite{zhang01,aspinice},
and supersolids \cite{tri1,tri2,tri3,tri4,moessner08,imp,tri_1stB,tri_sc,tri_1stA,tri_1stC,tocchio,sellmann}.
Frustrated lattices with spatial anisotropy
are in the center of attention
\cite{frank,cs1,org1,starykh,gan1,gan2,cs2,cs3,org2,isakov,white11A,white11B,org3,tocchio13,tocchio14,org4,meta,org5,org6,org7,becca16}
since a frustrating interchain
coupling allows to study effects of both a changing dimensionality and a tunable
frustration.

A straightforward frustrated geometry is provided by the triangular
lattice, which is realized by various types of materials,
ranging from antiferromagnets such as
Ba$_3$CoSb$_2$O$_9$ \cite{tri10,tri12,tri11}
and Cs$_2$CuCl$_{4-x}$Br$_x$ \cite{cs1,cs2,cs3} to
organic salts \cite{org1,org2,org3,org4,org5,org6,org7}.
For
antiferromagnetic $xy$-coupling with spatial anisotropy there has been a controversial
discussion on a possible
spin liquid phase \cite{cs1,cs2,cs3,org1,org2,org3,org4,org5,org6,org7,tocchio13,tocchio14,
becca16}
or an incommensurate phase \cite{white11A,white11B}.
For hard-core bosons in optical lattices the hopping parameter plays the role
of a ferromagnetic $xy$-coupling.
In this case, nearest neighbor interacting bosons are realizable, for instance, with magnetic erbium atoms \cite{ferlaino}.
On the triangular lattice \cite{sengstock,tri_nh} they have been predicted to show supersolid
behavior \cite{tri1,tri2,tri3,tri4,moessner08,imp,tri_1stA,tri_1stB,tri_1stC,tri_sc},
which is characterized
by two independent spontaneously broken symmetries -- U(1) and translation --
with corresponding
superfluid and density order.
For an anisotropic triangular lattice
the commensurate supersolid phase is found to be unstable \cite{gan1,gan2},
and the solid order turns out
to be incommensurate  \cite{isakov}.

\begin{figure*}[t]
\includegraphics[width=\textwidth,clip]{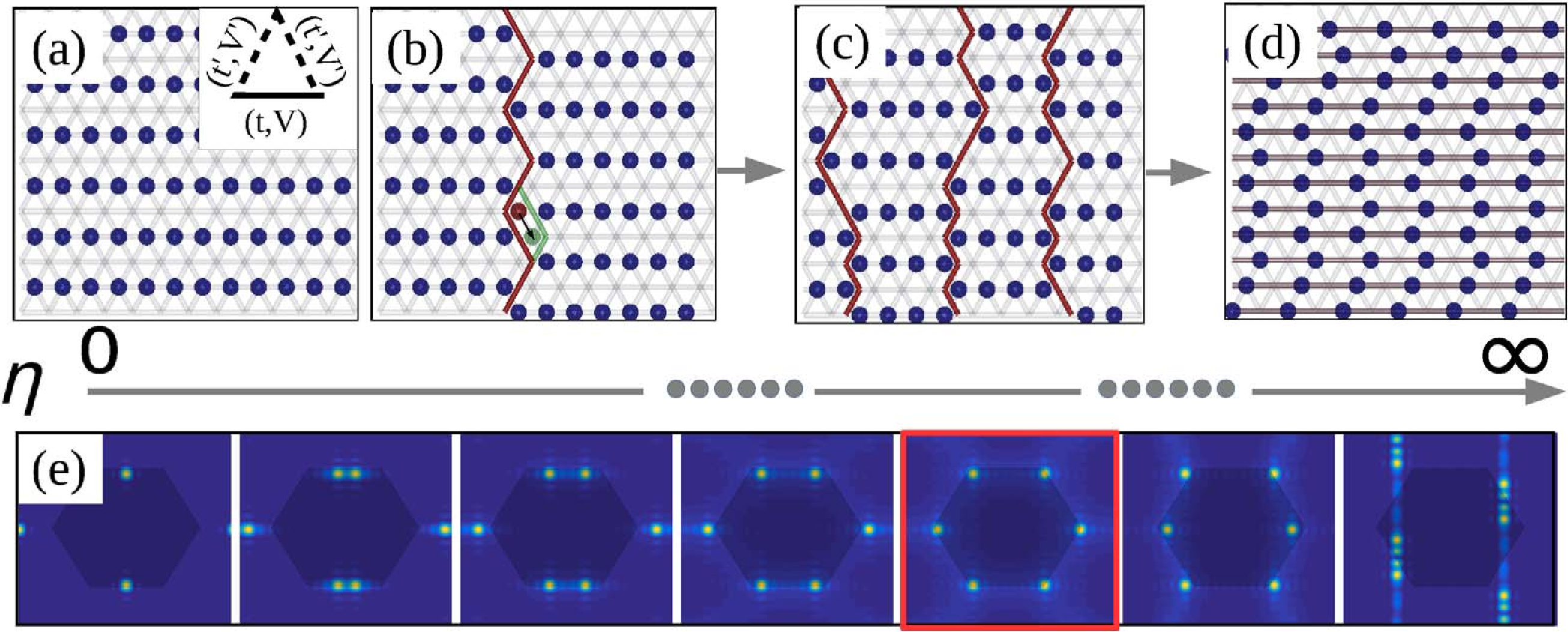}
\caption{(a) Configuration of checkered solid phase when $V'\gg V$
(inset: notation for hoppings and interactions);
(b) Single bosonic domain wall (red line) is excited for larger $\eta$.
Kinks can fluctuate by particle hopping to new shape (green);
(c) Multi-domain wall case;
(d) Decoupled chain phase when $V'\ll V$, thick lines indicate strong interactions and order
in $x$-direction;
(e) Structure factors of phases with different numbers of  domain walls
(from left to right: $N_D=$ 0, 2, 4, 6, 8, 10, and 12) for lattice size $L_x=L_y=12$ and pbc.
First Brillouin zone is indicated by shading and red
box marks commensurate supersolid ($\eta=1$).
\label{model}}
\end{figure*}

Supersolid phases with two independently broken order parameters were first discussed
for solid He \cite{supersolid2,supersolid1} and were more recently shown to exist
theoretically \cite{tri1,tri2,tri3,tri4,moessner08,imp,tri_1stA,tri_1stB,tri_1stC,tri_sc}
and experimentally \cite{hamburg,zurich} in optical lattices for ultra-cold gases.
Although high-temperature superconductors \cite{white01,vojta,metzner16} are
not often mentioned in this context,
the coexistence of superconductivity and
anti-ferromagnetic density order, so-called superstripes,
are the defining characteristics of an {\it incommensurate} supersolid.
Remarkably, the analogous
physical phenomenon  of incommensurate density order together with
a finite superfluid density can be observed in
a simple hard-core boson model.
In the following we present a quantitative analytical model for
this behavior in terms of topological defects in their
simplest form, namely an increasing number of domain walls.
Obviously the microscopic model of
high-temperature superconductors is quite different,
but the detailed understanding of the underlying mechanism via a spontaneous
appearance of domain walls \cite{vojta,metzner16,white01}
is a helpful unifying feature of these
many-body phenomena.

In this paper we analyze the quantum phase diagram of
hard-core bosons with anisotropic hopping $t,t' \geq 0$ and
 nearest neighbor interactions $V,V' \geq 0$
on a triangular lattice
\begin{eqnarray}
\nonumber
&\hat{H}&=\sumnear[
-t(\hat{b}_{i}^{\dag}\hat{b}_{j}+\mbox{h.c.})+
V (\hat{n}_{i}-1/2)(\hat{n}_{j}-1/2)
]\\
&+&\! \! \sumnn[
-t'(\hat{b}_{i}^{\dag}\hat{b}_{j}+\mbox{h.c.})+
V' (\hat{n}_{i}\!-\!1/2)(\hat{n}_{j}\!-\!1/2)
],
\label{eqHam}
\end{eqnarray}
where $\langle i,j\rangle_x$ and $\langle i,j\rangle_{\wedge}$ represent
the nearest neighbor sites on the horizontal bonds and the diagonal inter-chain bonds, respectively,
as shown in the inset of Fig.~\ref{model} (a).
According to Eq.~(\ref{eqHam})
we consider this model without additional
chemical potential, which
corresponds to half-filling,
where the most interesting physics is expected to occur
due to particle-hole symmetry. The anisotropy parameter is denoted by $\eta \equiv t/t'=V/V'$
and the number of lattice sites
in $x$- and $y$-directions are assumed to be equal $L=L_x=L_y$, yielding $N=L^2$.

In the strong-coupling limit $t=t'=0$, the hard-core boson model is equivalent
to the Ising model on the triangular lattice.
For the isotropic case $\eta=1$ any state, which fulfills the constraint of
one or two bosons per triangle, minimizes the energy. This leads to a
finite zero-temperature entropy of  $S/N=0.323~k_{\rm B}$ \cite{moessner1}, whose degeneracy is
removed for any $\eta\neq 1$.
When $\eta<1$, the bosons form a checkered order,  which alternates in the inter-chain direction as is shown in Fig.~\ref{model}(a).
This leads to distinct peaks
of the structure factor $S( \textbf{Q})=\langle |\mathop{\sum}_{k=1}^N n_k
e^{i \textbf{Q} \cdot\textbf{r}_k}|^2\rangle/N$
at the wave vectors $(\pm 2\pi,0)$ and $(0,\pm 2\pi/\sqrt{3})$, related by a
reciprocal lattice vector,
as is shown in the left panel of Fig.~\ref{model}(e).
When $\eta>1$, the particles form a quasi-density wave alternating in $x$-direction
with  order at
$\textbf{Q}=(\pm\pi,q_y)$
on each horizontal chain of the lattice according to Fig.~\ref{model}(d) and the right panel of Fig.~\ref{model}(e).
Here $q_y$ is arbitrary, i.e.~there is no order in the $y$-direction,
since the energy does not
change when all particles in one chain move together.
In this case the ground-state entropy is proportional to the number of chains $L_y$.

Finite hopping $t>0$ also removes the degeneracy, which leads to a finite range of
$\eta \neq 1$, where the ground state is dominated by quantum fluctuations.
For $\eta =1$ it is well known that the system supports a commensurate
supersolid phase \cite{tri1,tri2,tri3,tri4,moessner08,imp,tri_1stA,tri_1stB,tri_1stC,tri_sc}
at ordering wave vector $\textbf{Q}=(\pm 4\pi/3,0)$,
which is also stable in a range of non-zero chemical potential, i.e.~away from half-filling.
Remarkably, the supersolid can become incommensurate for $\eta\neq 1$ \cite{isakov}, which
will be described analytically in this work.

To this end we propose that an incommensurate supersolid can be modeled by topological excitations
of the checkered ordered phase.
In particular, the potential energy cost of inserting a domain wall in form of a phase shift as shown in
Fig.~\ref{model}(b) is given by $(V'-V) L_y/2=V' (1-\eta)L_y/2$, i.e.~it
is proportional to the number of
changed bonds in $y$-direction.
At the same time there is a gain in kinetic energy since each kink
can fluctuate
via particle hopping at the domain wall as indicated
by red and green sites in Fig.~\ref{model}(b). In fact the left/right direction
of the domain can be mapped to spin states (up/down) and
analyzing all possible cases of hopping it turns out that the kinetic energy is
described exactly by two independent $xy$-chains for each domain wall.
The $xy$-chain in turn is exactly solvable, including correlation functions, and
the energy is known to be $-2 L_y  t' / \pi$ \cite{zhang01}.
The effective energy for a finite density $\rho_D=N_D/L_x$ of domain walls is
therefore given by
\begin{eqnarray}
\label{energy}
E(\rho_D) = {L_x L_y}V'\rho_D\left[ \frac{1- \eta}{2} - \frac{2 t'}{\pi V'}
+  f (\rho_D) \right]  \, .
\end{eqnarray}
Here the last term accounts for an effective repulsive interaction energy between two neighboring domain walls separated
by a distance $1/\rho_D$.  The yet unknown function $f (\rho_D)$  must obey $f(\rho_D\!=\!0) =0$
and will be determined numerically below.
Using the condition $\partial_{\rho_D} E = 0$ in order to
extremize the domain wall energy (\ref{energy}) yields a relation between $\eta$ and $\rho_D$
\begin{eqnarray}
\label{eta}
\eta = 1 - \frac{4 t'}{V'\pi} + 2  \frac{\partial}{\partial \rho_D} \Big[ \rho_D f (\rho_D) \Big] \, .
\end{eqnarray}
The first domain wall appears when it becomes energetically favorable to spontaneously
allow fluctuations.
This onset of the phase transition can be determined from Eq.~(\ref{eta}) with
$\rho_D=0$ and $f=0$
resulting in a critical value $\eta_{c1}= 1 - 4 t' / V'\pi$,
which agrees well with numerical simulations, see below.
Note that additional domain walls at equal distance
will not immediately destroy the order completely.
Instead, each domain wall effectively removes half of a spatial density oscillation in a
finite system
without changing its wave length, so
that the positions of the structure factor peaks are shifted by $\pm \pi/L_x$
for each domain wall in the system.
This is the
microscopic origin of the observed incommensurable order, with  predicted
wave vectors $\textbf{Q} = \pi(\pm 2-\rho_D,0)$ and $\pi(\pm \rho_D,\pm 2/\sqrt{3})$
changing with $\rho_D$.
At the same time transport of bosons becomes possible along the
domain walls in $y$-direction,
leading to a corresponding anisotropic superfluid density which will be analyzed in
more detail below.

\begin{figure}[t]
\includegraphics[width=.5\textwidth]{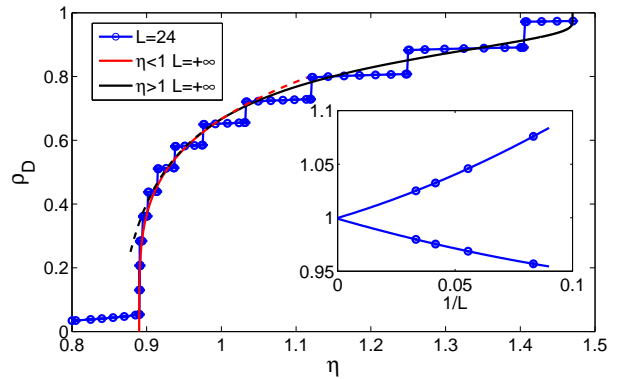}
\caption{Bosonic domain wall density (\ref{dw-density}) as
function of anisotropy parameter $\eta$
for $L=24$ with pbc
at $t'/V'=0.08$ and $\beta V_{\rm max}=400$ from QMC (dots)
compared to the analytical predictions from
Eqs.~(\ref{etafinal}) (red) and  (\ref{etafinal2}) (black), respectively.
Inset: Finite-size scaling of left and right points of the commensurate supersolid
plateau ($\rho_D=2/3$) with a second order polynomial fit.
\label{dm}}
\end{figure}

\begin{figure}[t]
\includegraphics[width=.5\textwidth]{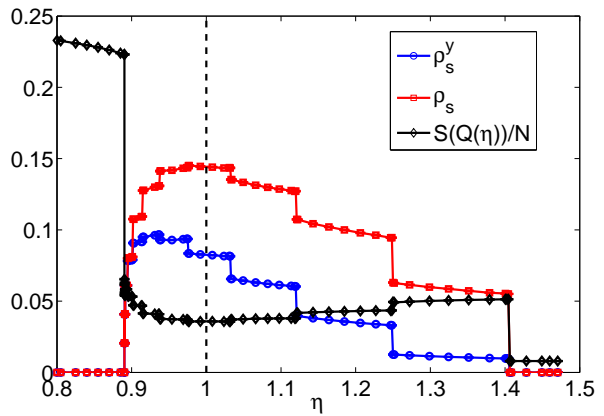}
\caption{Total superfluid density $\rho_s=\rho_s^x+\rho_s^y$, its value
in $y$-direction $\rho_s^y$, and
structure factor $S(\textbf{Q})/N$ at $t/V=0.08$,
$\beta V_{\rm max}=400$, and $L=24$.
\label{sf}}
\end{figure}

For the numerical results we implemented a Quantum Monte Carlo (QMC) code using the
Stochastic Cluster Series Expansion algorithm
\cite{sseA,sseB,clustersse}, which is further
discussed in the Appendix \cite{sm}. Since the bosonic hopping is positive,
there
is no kinetic frustration and the model does not suffer from the minus sign problem.
We use $5\times10^5$ thermalization steps and $10^6$
measuring steps on lattices up to length $L=24$ with periodic boundary conditions (pbc) in
both directions.
Since topological quantum numbers, such as domain walls, are difficult to change
with ordinary QMC updates, we developed an extension of the parallel tempering method
\cite{sm,tempering}.
All numerical data in Figs.~\ref{dm} and \ref{sf} are plotted with error bars
which are too small to be distinguishable.
The inverse temperature $\beta$
is measured in units of the larger repulsive energy
$V_{\rm max}=\max(V,V')$.
The transition from the supersolid phase to a uniform superfluid phase occurs at
$t'/V'\approx0.11$ \cite{tri1}.  In our simulations we therefore focus on the choice of
$t/V=t'/V' = 0.08$, which is in the
center of the supersolid phase we are interested in.  Results for other values of $t'/V'$
are discussed in the Appendix \cite{sm}.

In order to determine the domain wall density we define the operator
\begin{eqnarray}
\rho_D=\mathop{\sum}_{i_y=1}^{L_y}\mathop{\sum}_{i_x=1}^{L_x}\frac{n_{(i_x,i_y)}\bar{n}_{(i_x+1,i_y)}+n_{(i_x+1,i_y)}\bar{n}_{(i_x,i_y)}}{L_xL_y}\,,
\label{dw-density}
\end{eqnarray}
where $\bar{n}=1-n$.  For hard core bosons $n_{(i_x,i_y)}\bar{n}_{(i_x+1,i_y)}+n_{(i_x+1,i_y)}\bar{n}_{(i_x,i_y)}$
corresponds to
$1-\delta_{n_{(i_x,i_y)},\bar{n}_{(i_x+1,i_y)}}$, so $\rho_D$ effectively counts the
number of density changes along the $x$-direction.
Note that domain walls have to be created in pairs because of pbc, so that
$N_D=\rho_D L_x$ should be an even integer.
The QMC results in Fig.~\ref{dm} clearly show
plateaus of quantized domain wall numbers with
discrete jumps as a function of $\eta$.
This is remarkable since the densities fluctuate strongly in the QMC simulations,
but the correlation function in Eq.~(\ref{dw-density}) yields robust discrete
quantum numbers. The expected integer values are only slightly
renormalized by quantum fluctuations, which are
further suppressed for smaller $t'/V'$ \cite{sm}.
The phase transition occurs at $\eta_{c1} \approx 0.89$
for $t'/V'=0.08$ in good agreement with the analysis above.
The plateaus and the jumps between them become more and more continuous
for increasing lengths.  A finite-size scaling of the left and right ends of
the commensurate plateau at $\rho_D=2/3$ is shown in the inset of Fig.~\ref{dm},
which clearly
demonstrates that a continuous function $\rho_D(\eta)$ is approached in the
thermodynamic limit.  Also the jump to the first plateau and the change in
the superfluid order parameter vanishes in the thermodynamic limit at the critical
point $\eta_{c1}$ \cite{sm}, which is the hallmark of a second-order phase transition.
The interaction energy $f(\rho_D)$ can be analyzed from QMC results
with the help of Eq.~(\ref{energy}),
which yields a near perfect agreement with
a powerlaw behavior $f(\rho_D) \propto \rho_D^4$ \cite{sm}, i.e.~a 1/distance$^4$-law.
The proportionality constant is fixed by imposing the
communicability condition $\rho_D(\eta\!=\!1)=2/3$, which together with Eq.~(\ref{eta}) yields
a prediction for the behavior in the thermodynamic limit
\begin{eqnarray}
\label{etafinal}
\rho_D(\eta) = \frac 2 3
\left({\frac{\eta-\eta_{c1}}{1-\eta_{c1}}}\right)^{1/4}.
\end{eqnarray}
This analytic results is shown as a red line in Fig.~\ref{dm} with very good agreement
in the region of dilute domain walls, i.e.~$\rho_D<2/3$ and $\eta_{c1}<\eta <1$.

At full saturation $\rho_D=1$ a second transition to
the decoupled chain phase occurs at $\eta_{c2}$.
This phase has an independent alternating density order in the ground state.
The basic excitation is again a domain wall, but this time for each chain separately in the
form of equal neighboring densities $n_{i,j}=n_{i,j+1}$, i.e.~a phase shift in
the alternating order along $x$-direction.
Such a neighboring density pair can be transported ballistically along the
entire chain, which yields a large superfluid density along the $x$-direction and again
a shift in the ordering wave vector.  A single pair has a potential energy cost of
$(V-V')/2$ and yields a kinetic energy gain in the $q_x=0$ state of $-2 t$, resulting in
a critical density of $\eta_{c2}= 1/(1 - 4 t'/V')$. This yields for $t'/V'=0.08$ the value $\eta_{c2} \approx 1.47$
in excellent agreement with our numerical results. We can also define the
interaction energy between two neighboring density pairs $g (\rho_D)$, and
determine it along similar lines as above. A fit to numerical results yields for
$g (\rho_D)$ an
exponential decay \cite{sm} which gives a domain wall function
\begin{eqnarray}
\label{etafinal2}
\rho_D(\eta) = 1+ \frac{1}{3} \left[ 1+W_{-1} \left( -\frac{2(\eta_{c2}-\eta)}{e^2(\eta_{c2}-1)} \right) \right]^{-1}
\end{eqnarray}
in the limit of few density pairs, i.e.~relatively close to saturation $2/3 \alt \rho_D < 1$.
 Here $W_{-1}$ represents the branch $-1$ of the Lambert $W$ function and is plotted in Fig.~2 as a black line for $t'/V'=0.08$ and
$1 \leq \eta < \eta_{c2}$.


Further evidence for the existence of domain walls comes from the
corresponding structure factor $S(\textbf{Q})/N$, which is shown in
Fig.~\ref{model}(e) at one parameter point for each domain wall number.
In the incommensurate supersolid the peak positions shift with anisotropy, which reflects the
change from  checkered order to independent chains.
As predicted by the domain wall theory,
the peak positions are directly related to the respective number of domain walls \cite{sm}.
All these observations strongly support the quantum nature of domain walls, and rule out
classical explanations of incommensurate order,
such as  a continuous spiral rotation of the spin \cite{zheng}.

In the supersolid phase both the translational symmetry and the U(1)
gauge symmetry are broken. Therefore it is also interesting to analyze the
corresponding U(1) order parameter, namely the
total superfluid
density $\rho_s=\rho_s^x+\rho_s^y$ and its components $\rho_s^{x(y)}=W_{x(y)}^2/(4\beta t)$, where $W_{x(y)}$ stands for
the winding number in $x(y)$-direction \cite{winding}. According to Fig.~\ref{sf}, the
total superfluid density $\rho_s$ behaves opposite to the structure factor, which
indicates that the two order parameters are competing \cite{imp}.
The total superfluid density
is reduced in both anisotropic limits and increases for a decreasing anisotropy.
In general the superfluid density is an anisotropic tensor \cite{ueda,sf}, which
in fact reflects the properties of the domain walls
presented above:
For $\eta<1$ the superfluidity along the $y$-direction dominates, which is
caused by hopping of domain wall kinks, increasing with the number of domain walls.
The maximum of the total superfluid density
occurs at $\eta=1$, where fluctuations in both directions are equally possible.
For $\eta > 1$ the superfluidity is caused by neighboring density pairs, predominantly in the
x-direction. Interestingly, we observe strong jumps in $\rho_s^y$ at $\eta = 1.25$
 and $\eta = 1.1$
corresponding to the occurance of the second and third density pair in a finite system.
This can only be explained by correlations of the density pairs perpendicular to the
chains, which signals the buildup of domain walls discussed above for $\eta<1$.  The jumps
are a finite-size effect, however, so the behavior becomes continuous in the thermodynamic
limit.

In the context of superfluidity it is interesting
to remember that also in the striped phase of
high-temperature superconductors domain walls and superconductivity
coexist \cite{vojta,metzner16,white01}.
We have seen that in our bosonic model both domain walls and neighboring density
pairs naturally emerge and cause superfluidity where the fluctuations are largest.
Even though the microscopic pairing mechanism is more involved for
superconductors, our results
suggest that it is related to the coherent motion of neighboring domain walls and
an effective reduction to one dimension.

In summary, we analyzed the quantum phase diagram of the
extended anisotropic Bose-Hubbard model on the triangular lattice.
Due to frustration a non-trivial incommensurate supersolid phase appears, which
can be well described analytically by topological defects in the form of
domain walls.  For small $\eta$ the domain walls
along the $y$-direction
are described by the exactly solvable $xy$-chain model
together with a 1/distance$^4$-interaction between them.  For large $\eta$ an
independent description in terms of neighboring density pairs with an exponential
interaction is possible.  The numerical results for the
phase transition lines, the domain wall density, the incommensurate ordering
wave vectors, and the superfluid density agree with this theory.
In the low density region it will, in principle, be possible to use a more
detailed analysis of the $xy$-model, if  higher-order correlation functions,
finite temperature behavior, or dynamical properties need to be calculated.
For larger domain wall densities a coherent pairing mechanism
causes a
dominant superfluid density in the $x$-direction,
which is due to the movement of
neighboring density pairs.
This mechanism may be related to pairing in
the striped phases in high temperature
superconductors, where superconductivity coexists with incommensurate order and domain walls.


\begin{acknowledgments}
We are thankful for useful discussions with Chisa Hotta, Ying Jiang, Frank Pollmann,  and Yue Yu. This work was supported
by the Chinese Academy of Sciences
via the Open Project Program of the State Key Laboratory of
Theoretical Physics, by the Nachwuchsring of the TU Kaiserslautern, 
and  by the German Research Foundation (DFG)
via the Collaborative Research Centers
SFB/TR49, SFB/TR173, and SFB/TR185.
The authors gratefully acknowledge the computing time granted 
by the John von
Neumann Institute for Computing (NIC) on the supercomputer JURECA at
J\"ulich Supercomputing Centre (JSC) and by the 
Allianz f\"{u}r Hochleistungsrechnen Rheinland-Pfalz (AHRP),
\end{acknowledgments}

\bibliographystyle{apsrev}

\section{Appendix}

In the Appendix additional information on the
performed Quantum Monte Carlo (QMC) simulations are given. 
Further results as a function of hopping strength and 
system size are shown.  In addition to periodic boundary conditions, 
cylindrical boundary conditions also are considered.  Finally, details
on the derivation of the effective 
interaction energy between the domain walls are presented.

\subsection{Quantum Monte Carlo method}

For the numerical quantum Monte Carlo (QMC) simulations we implemented 
the cluster stochastic series expansion method \cite{sseA,sseB,clustersse} 
by taking into account three sites as the update unit,  
as it can help increasing the ergodicity \cite{clustersse}. 
It is well known that QMC may have an infamous minus sign problem
if an exchange of two fermions is possible, or if a kinetic frustration occurs, 
i.e.~the overall probability of an off-diagonal exchange along a closed loop is negative.
Since we are dealing with bosons with positive hopping our model 
is without sign problem, despite the triangular geometry.

However, another typical QMC problem of trapping in a local minimum is indeed
of concern for our studies.  In particular, 
the domain walls in the incommensurate supersolid phase 
represent very robust topological defects, so the system may get trapped at a 
fixed domain wall
number, which is difficult to change with a small hopping parameter even 
using the loop update. In addition, because 
the wave functions with different domain wall numbers have 
nearly no overlap, also ordinary parallel tempering \cite{tempering}
 does not show much improvement.
In order to solve this local minimum problem, a further extension 
of the parallel tempering method has been developed. To this end we perform 
the normal QMC simulation in the whole parameter region first and store 
the configuration at each point. 
Then, for each parameter point, we use the outcome of neighboring 
parameter points as 
the initial  configuration and launch additional simulations. 
When they are finished, we choose the simulation with the lowest average energy 
as the right one since at low temperatures we are effectively in the ground state limit. 
We find that without this change of initial configurations the 
resulting energy curve is discontinuous as a function of parameters, but it
becomes continuous after the swap. 
Only for larger hopping parameters, 
such as $t/V=0.1$, the ordinary parallel tempering with loop update can overcome 
the local minimum.

\begin{figure}[h]
\includegraphics[width=.45\textwidth]{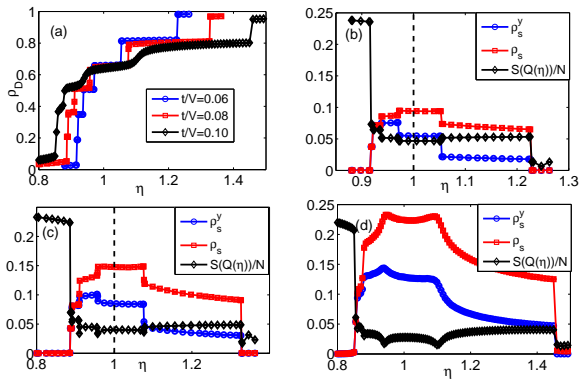}
\caption{(a) Bosonic domain wall density as
function of anisotropy parameter $\eta$
at $t/V=0.06,0.08,0.10$.  (b-d): 
Total superfluid density $\rho_s$, its value
in $y$-direction $\rho_s^y$, and
structure factor $S(\textbf{Q})/N$ at $t/V=0.06$ (b), $t/V=0.08$ (c), $t/V=0.10$ (d).
All QMC results have been obtained for $\beta V_{\rm max}=200$ and $L=12$
with periodic boundary conditions.
\label{L12_rho}}
\end{figure}

\subsection{Phase Transitions}

The domain wall picture works well in the strong-coupling limit. 
In fact up to first-order perturbation theory in the hopping parameter, 
the ground states with different domain 
walls have no overlap, so there are no corrections to the quantized values of the
domain wall numbers.   In the following we illustrate the 
effect of changing hopping close to (but below)
the superfluid transition at $t'/V'\approx0.11$ \cite{tri1}.
Figure \ref{L12_rho} (a) shows that the steps are very clearly visible in the 
strong-coupling region at $t'/V'=0.06$, but relaxation of the plateaus is observed 
when the hopping is increased to $t'/V'=0.1$. 
The same is true for the step-like structure for all order 
parameters in Fig.~\ref{L12_rho}(b-d).
For larger hopping $t'/V'=0.1$ and near the ends of the plateaus,
we observe maxima in the superfluid density and minima in the structure factor
due to increased fluctuations in Fig.~\ref{L12_rho}(d), but the qualitative signature of
domain walls always remains visible.

\begin{figure}[t]
\includegraphics[width=.45\textwidth]{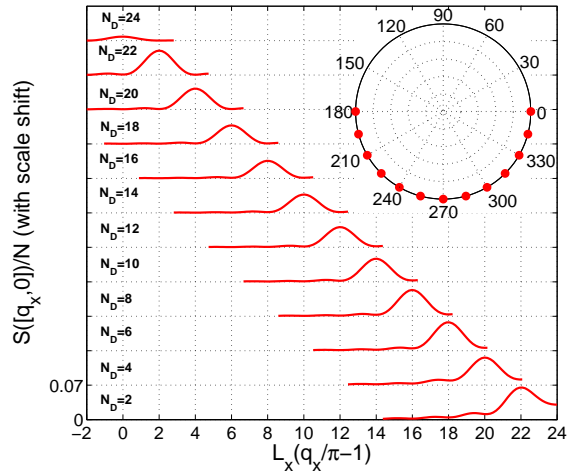}
\caption{Structure factor $S(\textbf{Q})/N$ as 
function of $q_x$ at $q_y=0$ for different domain wall
densities (shifted relative to each other) at $t/V=0.08$, $L=24$, and $\beta V_{\max}=400$.
Inset: Maximum position (red dot) of $S(\textbf{Q})/N$ in $q_x$-direction for $q_y=0$ using values of $\eta$
corresponding to different domain wall numbers $N_D$.
\label{sfqkx}}
\end{figure}

One of the main results in the paper is that 
each plateau of the domain wall density is closely related to a corresponding maximum 
of the structure factor. In order to make this connection 
clearer, we plot the structure factor $S(\textbf{Q})/N$ 
in $q_x$-direction for $q_y=0$ using values of $\eta$
in the center of the corresponding domain wall plateau in Fig.~\ref{sfqkx}. 
The positions of the peaks match well with the analytic prediction $N_D=L_x(2-q_x/\pi)$.

\subsection{Finite-size scaling}
 
\begin{figure}[t]
\includegraphics[width=.45\textwidth]{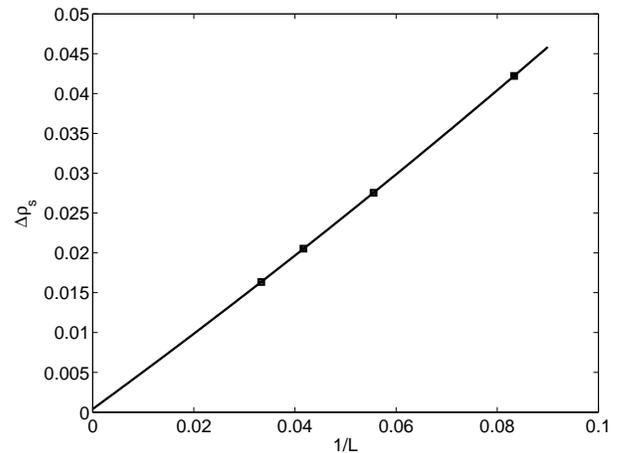}
\caption{Finite-size scaling of superfluid density difference between the two phases at $\eta_{c1}$
for $t/V=0.08$, $\beta V_{\max}=200$, and $L/12$.
\label{eta0_ffs_rhos}}
\end{figure}

%
In the paper we demonstrated that the quantization of the domain wall density 
becomes continuous in the thermodynamic limit. This implies that the shift of the 
structure factor
is also continuous at the phase transition, indicating a second-order phase transition.
To give additional support we present the finite-size scaling of
the superfluid density difference between the two phases at $\eta_{c1}$
in Fig.~\ref{eta0_ffs_rhos}.  Clearly, the second-order polynomial fit shows that 
the jump also vanishes in the thermodynamic limit 
for this order 
parameter, which confirms that the phase transition is of second order.

\subsection{Cylindrical boundary}

\begin{figure}[t]
\includegraphics[width=.45\textwidth]{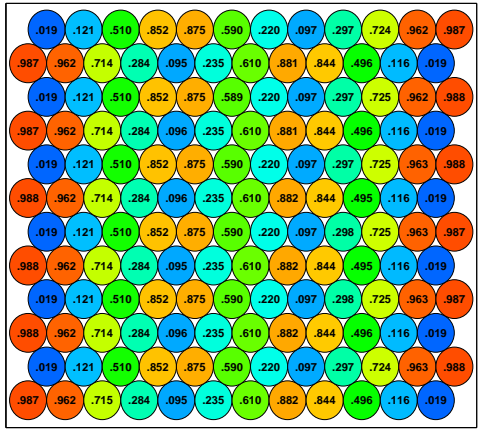}
\caption{Average bosonic density distribution at $t/V=0.08$, $\beta V_{\max}=200$, $\eta=0.9$, and $L = 12$ with $y$-cbc.
\label{rhos}}
\end{figure}

Whereas so far we have chosen periodic boundary conditions, we investigate now cylindrical 
boundary conditions ($y$-cbc) with open ends in $x$-direction. In that case single 
domain walls can be created, so the number of domain walls $N_D=\rho_D L$ can be both 
even or odd integers, which is indeed seen in the simulations. Furthermore,
Fig.~\ref{rhos} shows that
the fluctuating domain walls are clearly visible in terms of regions of average
half-filling, which separate regions of checkered order, the latter being particularly stable near the
edges. All these observations strongly support the quantum nature of domain walls, and rule out
classical explanations of incommensurate order,
such as  a continuous spiral rotation of the spin \cite{zheng}.

\begin{figure}[t]
\includegraphics[width=.45\textwidth]{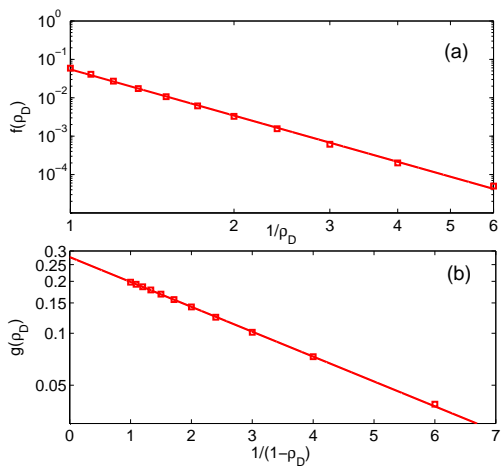}
\caption{Fitting of interaction energies between two neighboring (a) domain walls $f(\rho_D)$, (b) density pairs $g(\rho_D)$ for 
$t/V=0.08$, $\beta V_{\max}=400$, and $L = 24$. 
\label{lt}}
\vspace{-3.5mm}
\end{figure}

\subsection{Effective interaction energies}

In this section, we discuss a phenomenological model 
how to determine the effective interaction energies between two neighboring domain walls for $\eta<1$ and density pairs for $\eta>1$ from QMC data. 
At first, we discuss the dilute domain wall case $\eta<1$. 
The total energy of $N_D$ domain walls is given by
\vspace{-1.5mm}
\begin{eqnarray}
E(N_D) =  N_D L_y \left[ \frac{V' - V}{2} - \frac{2}{\pi} t'  + f\left(\frac{N_D}{L_x}\right) V' \right] \, ,
\end{eqnarray}
as discussed in the Letter. 
Here the first term corresponds to the potential energy with $V=\eta V'$, the second term denotes the kinetic energy.
The third term models the repulsive interaction energy between the domain walls, where the function 
$f(N_D/L_x)$ takes into
account the dependence on the distance $L_x/N_D$ between two domain walls. Thus, we arrive at
\vspace{-1.5mm}
\begin{eqnarray}
E(N_D) = V' N_D L_y \left[ \frac{1 - \eta}{2}  - \frac{2}{\pi} t'/V' +f\left(\frac{N_D}{L_x}\right) \right] \, .
\end{eqnarray}
Due to periodic boundary conditions the number of domain walls $N_D$ changes by multiples of 2. In order to analyze the respective transitions from $N_D = 2M-2$
to $N_D = 2M$ with $M=1, \ldots, M_{\rm max}=L_x/2$ we have to evaluate the conditions
\vspace{-1.5mm}
\begin{eqnarray}
E(N_D=2M-2) =E(N_D=2M) \, .
\end{eqnarray}
This yields the position of the jumps between the plateaus 
\begin{eqnarray}
\label{eta2}
\nonumber
\eta_M &=& 1 - \frac{4}{\pi}t'/V' +2M f\left(\frac{2M}{L_x} \right) \\
&&  - (2M-2)f\left(\frac{2M-2}{L_x} \right) \, .
\end{eqnarray}
By formally identifying $\eta_0=\eta_{c1}=1-4t'/V' \pi$, the respective
jump points with different domain wall numbers read explicitly
\vspace{-3.5mm}
\begin{eqnarray}
\label{etab}
\nonumber
\eta_1&=&\eta_{c1}+2f(2/L_x)\\
\nonumber
\eta_2&=&\eta_{c1}+4f(4/L_x)-2f(2/L_x)\\
\nonumber
\eta_3&=&\eta_{c1}+6f(6/L_x)-4f(4/L_x)\\
& \vdots &
\end{eqnarray}
Thus, we deduce from (\ref{etab}) that the effective interaction energy can be reconstructed from the values at which the jumps occur according to
\vspace{-3.5mm}
\begin{eqnarray}
\label{efe}
f\left(\frac{2M}{L_x}\right)=\sum_{i=1}^M \frac{\eta_i}{2M}-\frac{\eta_0}{2}\, .
\end{eqnarray}
Using the jump points from QMC simulations (c.f.~Fig.~2 in the Letter) we determine via 
Eq.~(\ref{efe}) the effective interaction energy between
two domain walls $f(\rho_D)$ for a finite density $\rho_D=N_D / L_x$. 
The results are shown in Fig.~\ref{lt} (a) on a double logarithmic scale
 and clearly indicate a simple power law
\vspace{-3.5mm}
\begin{eqnarray}
f(\rho_D) \sim \rho_D^\alpha  \, , \label{f}
\end{eqnarray}
with $\alpha= 4 \pm 0.1$.
The same strategy turns out to be also applicable in the decoupled chain region with
neighboring density pair excitations, 
i.e.~$1 < \eta$. The only difference
is that now $\eta_0=\eta_{c2}$ and the effective interaction energy 
between two neighboring density pairs is denoted by another function 
$g(\rho_D)$ with $g(\rho_D\!=\!1)=0$. The corresponding fit in Fig.~\ref{lt} (b) reveals 
that the effective interaction energy $g(\rho_D)$ is given by an exponential
\vspace{-3.5mm}
\begin{eqnarray}
g(\rho_D) \sim \exp \left( - \frac{1}{\gamma(1-\rho_D)} \right)  \, , \label{g}
\end{eqnarray}
with $\gamma = 3 \pm 0.05$.  The results in Eqs.~(\ref{f}) and (\ref{g}) are used in the 
Letter to 
derive a relation between $\eta$ and $\rho_D$ in thermodynamic limit.

\end{document}